\documentclass[%
 reprint,
amsmath,amssymb,aps,
pra,
]{revtex4-2}

\usepackage{graphicx}
\usepackage{dcolumn}%
\usepackage{bm}

\begin{document}
\title{Gravity induced CP violation in mesons \\
mixing, decay and interference experiments}
\author{J-M Rax}
\email{jean-marcel.rax@universite-paris-saclay.fr}
\affiliation{IJCLab-UMR9012-IN2P3 \\Universit\'{e} de Paris-Saclay\\  
 Facult\'{e} des Sciences d'Orsay\\
91405 Orsay France}
\date{\today}

\begin{abstract}
    The impact of earth's gravity on neutral mesons dynamics is analyzed. 
    The main effect of a Newtonian
    potential is to couple the strangeness and bottomness oscillations with the
    quarks zitterbewegung oscillations. This coupling is responsible for the
    observed CP violations in the three types of experiments analyzed here: ({\it i})
    indirect violation in the mixing, ({\it ii}) direct violation in the decay
    to one final state and ({\it iii}) violation in interference between decays
    with and without mixing. The three violation parameters associated with
    these experiments are predicted in agreement with the experimental data. The
    amplitude of the violation is linear with respect to the strength of gravity
    so that this new mechanism allows to consider matter dominated cosmological
    evolutions providing the observed baryon asymmetry of the universe.
\end{abstract}

\maketitle

\section{Introduction}

Since the first observation of long-lived kaons decays into pairs of charged
pions, reported sixty years ago by Christenson, Cronin, Fitch and Turlay 
\cite{1}, many complementary observables associated with flavored neutral
mesons CP violation (CPV) have been identified, measured and interpreted.
The canonical framework of interpretation is the standard model (SM) through
the adjustment between the Kobayashi-Maskawa (KM) \cite{2} complex phase and
the experimental values of the violation parameters. In this study, we focus
on the most documented and clearest experimental evidence of CPV and we
demonstrate that gravity induced CPV provides a pertinent framework to
interpret these  experiments and to predict the violation parameters, as a
function of earth's gravity, in agreement with the experimental data. As a
consequence, far from any massive object, i.e. in a flat Lorentzian
space-time, the Cabibbo-Kobayashi-Maskawa (CKM) \cite{2,3} matrix must be
considered free from any CPV phase as CPV effects are just gravity induced
near massive objects like earth.

Among the measured CPV observables, three types ({\it i}, {\it ii} and 
{\it iii}) of effects will be considered here: ({\it i}) indirect CPV
in the mixing which has been observed with neutral kaons $K^{0}/\overline{K}^{0}$, 
this CPV is described by the parameter $\mathop{\rm Re} \varepsilon $ \cite{4}; ({\it ii}) 
direct CPV in decays into one final state which has also been observed
in neutral kaons decays and is characterized by the parameter $\mathop{\rm
Re} \varepsilon ^{\prime }/\varepsilon $ \cite{5}; ({\it iii}) CPV in
interference between decays with and without mixing, which has been observed
in $B^{0}/\overline{B}^{0}$ decays and is described by the angle $\beta $ 
\cite{5}.

Beside these ({\it i}, {\it ii} and {\it iii}) types of CPV, a
forth additional experimental evidence of CPV must be considered: ({\it iv}) 
the observed dominance of baryons over antibaryons in our universe as CPV
is one of the necessary condition to build cosmological evolution models
compatible with this baryon asymmetry \cite{6}.

Despite its success to provide a framework to interpret earth based
experiments such as ({\it i}, {\it ii} and 
{\it iii}), the KM
mechanism, incorporated into cosmological evolution models, fails, by many
orders of magnitude, to account for this ({\it iv}) major CPV evidence.
To explain how our matter-dominated universe emerged during its early
evolution we need to identify a CPV mechanism different from the KM one.
Beside its potential to predict the measured parameters associated with
types ({\it i}, {\it ii} and 
{\it iii}) CPV experiments, the new
gravity induced CPV mechanism opens very interesting perspectives to set up
cosmological models with asymmetric baryogenesis compatible with the present
state of our universe as the amplitude of CPV appears here to be
proportional to the strength of gravity. During the early stages of
evolution of the universe, gravity/curvature was far more larger than on
earth today and gravity induced CPV identified and described here, which is
a linear function of the gravitational field, opens an avenue to resolve the
present contradiction between the very small value of the KM mechanism and
the very large CPV needed to build a pertinent model of our matter-dominated
universe. To summarize, gravity induced CPV, not only explain ({\it i}, {\it ii} and 
{\it iii}) CPV effects and predict observables such as $%
\varepsilon $, $\varepsilon ^{\prime }$ and $\beta $, but it also renews, in
depth, the baryons asymmetry ({\it iv}) cosmological issue.

In this study, we demonstrate that a small secular coupling, induced by
earth's gravity, between fast {\it quarks zitterbewegung oscillations} at
the velocity of light inside the mesons and strangeness oscillations $\Delta
S=2$, or bottomness oscillations $\Delta B^{\prime }=2$, provides both a
qualitative explanation of CPV and a quantitative prediction of the CPV
parameters $\varepsilon $, $\varepsilon ^{\prime }$ and $\beta $ in
agreement with the experimental measurements.

The new interpretation of CPV experiments presented below is based on a
careful analysis of the impact of earth gravity on the dynamics of
strangeness and bottomness oscillations. To do so we use the effective
Hamiltonian of Lee, Oehme and Yang (LOY) \cite{7,8}, completed here with
Newtonian gravity. Neutral mesons oscillations such as $K^{0}%
\rightleftharpoons \overline{K}^{0}$ $\ $ and $B^{0}\rightleftharpoons 
\overline{B}^{0}$ are very low energy oscillations ($10^{-6}-10^{-4}$ eV).
The typical earth's gravity coupling parameter $\hbar g/c\sim 10^{-23}$ eV ($%
g$ is the acceleration due to gravity on earth) is very small with respect
to the various energy scales involved in neutral mesons oscillations. Given
the smallness of these parameters, there are no needs to rely on quantum
field theory and the usual two states LOY model offers the pertinent
framework to describe the interplay between two low energy quantum
oscillations: quarks zitterbewegung vertical oscillations at the velocity of
light in a gravitational field on the one hand and the strangeness
oscillations $\left( \Delta S=2\right) $, or\ bottomness oscillations $%
\left( \Delta B^{\prime }=2\right) $, on the other hand.

The three types of CPV experimental evidences are usually analyzed under the
assumption of CPT conservation. The CPT theorem is demonstrated within the
framework of three hypothesis: Lorentz group invariance, spin-statistics
relations and local field theory. In the rest frame of a meson interacting
with a massive spherical object like earth the first hypothesis is not
satisfied. Thus, when earth influence is considered, we must not be
surprised that CPT theorem, apparently, will no longer holds. Within the
framework of a gravity induced CPV mechanism earth's gravity is described as
an external field and the evolution of a meson state $\left| M\right\rangle $
alone, as a linear superposition of two flavor eigenstates $\left|
M^{0}\right\rangle $/$\left| \overline{M}^{0}\right\rangle $, does not
provide the complete picture of the dynamical system and so can not be
considered as a good candidate displaying CPT invariance. But CPT might be
restored for the global three bodies $\left( M^{0}/\overline{M}^{0}/\oplus
\right) $ evolution of the state $\left| M\oplus \right\rangle $ describing
both the meson-antimeson pair and earth. In this study we consider only the
evolution of $\left| M\right\rangle $ and earth's effect is described as an
external static field so that CPT will appear to be violated because of \
this restricted two bodies $\left( M^{0}/\overline{M}^{0}\right) $ model of
the system.

The study presented below complements a previous study based on two coupled
Klein-Gordon equations describing $K^{0}/\overline{K}^{0}$ evolution on a
Schwarzschild metric \cite{9}, rather than a Newtonian framework with \ two
coupled Schr\"{o}dinger equations used here. The results on $K^{0}/\overline{%
K}^{0}$ dynamics given by the Newtonian model, presented below, are similar
to those of this previous Einsteinian model \cite{9}, these results are thus
model independent. Moreover, with gravity induced CPV there is no T
violation at the microscopic level and, for example in the $K^{0}/\overline{K%
}^{0}$ case, the observed T violation stems from the irreversible decay of
the short-lived kaons $K_{S}$ continuously regenerated from the long-lived
one $K_{L}$ by the gravity induced coupling.

This paper is organized as follows, in the next section we briefly review
the LOY model without CPV. In section III we review the usual modifications
to $K^{0}/\overline{K}^{0}$ and $B^{0}/\overline{B}^{0}$ mass eigenstates
needed to accommodate CPV experimental results. The impact of earth's
gravity is considered in section IV where, to describe neutral mesons
oscillations $M^{0}\rightleftharpoons \overline{M}^{0}$on earth, the CP
conserving LOY model, presented in section II, is completed with a Newtonian
gravity term. We carefully analyze the nature and the impact of this
additional term and discover that it contains the zitterbewegung motion at
the velocity of light of the quarks inside the meson. The study of type (%
\textit{i}), (\textit{ii}) and (\textit{iii}) gravity induced CPV are
developed in sections V, VI and VII. We consider specifically type (\textit{i%
}) and (\textit{ii}) CPV for $K^{0}/\overline{K}^{0}\sim \left( d\overline{s}%
\right) /\left( \overline{d}s\right) $ and type (\textit{iii}) CPV for $%
B^{0}/\overline{B}^{0}\sim \left( d\overline{b}\right) /\left( \overline{d}%
b\right) $. Section VIII provides a brief analysis of others, $D^{0}/%
\overline{D}^{0}$ and $B_{s}^{0}/\overline{B_{s}}^{0}$, neutral mesons\ and
gives our conclusions. In sections II and IV $M^{0}/\overline{M}^{0}$ will
stand for $K^{0}/\overline{K}^{0}$ or $B^{0}/\overline{B}^{0}$.

\section{Mass and CP eigenstates without CPV}

Consider a generic neutral meson pair $M^{0}/\overline{M}^{0}$, either $%
K^{0}/\overline{K}^{0}$ or $B^{0}/\overline{B}^{0}$. The meson state $\left|
M\left( \tau \right) \right\rangle $ is a linear superposition of the flavor
eigenstates $\left| M^{0}\right\rangle $ and $\left| \overline{M}%
^{0}\right\rangle $ and the amplitudes $\left( a,b\right) $ of this
superposition are functions of the meson proper time $\tau $. This state is
also coupled to a set of final states $\left| f,\mathbf{p,...}\right\rangle $%
,\ with momentum $\mathbf{p}$ in the $M^{0}/\overline{M}^{0}$ meson rest
frame, described by the amplitudes $w_{f}$,

\begin{equation}
\left| M\left( \tau \right) \right\rangle =a\left( \tau \right) \left|
M^{0}\right\rangle +b\left( \tau \right) \left| \overline{M}%
^{0}\right\rangle +\sum_{f}w_{f}\left( \tau \right) \left| f\right\rangle 
\text{,}  \label{a1}
\end{equation}
where $\left\langle M^{0}\right. \left| M^{0}\right\rangle $ $=$ $%
\left\langle \overline{M}^{0}\right. \left| \overline{M}^{0}\right\rangle $ $%
=$ $1$ and $\left\langle \overline{M}^{0}\right. \left| M^{0}\right\rangle
=0 $.

The Weisskopf-Wigner (WW) approximation \cite{10} is used to describe the
coupling to the final states $\left| f\right\rangle $ as an irreversible
decay. Within the framework of this usual approximation we introduce a
non-Hermitian decay operator $j\widehat{\gamma }$ capturing the effects of
the $w_{f}$ amplitudes and describing $M\rightarrow f$ \ transitions as an
irreversible decay process. It is to be noted that, as the possibilities of $%
f\rightarrow $ $M$ transitions are neglected by this approximation, the use
of $j\widehat{\gamma }$ is thus the source of a T violation which must not
be attributed to fundamental interactions but to the WW model.

The time evolution of $\left| M\left( \tau \right) \right\rangle $ can thus
be restricted to a two states Hilbert sub-space: $\left[ \left|
M^{0}\right\rangle ,\left| \overline{M}^{0}\right\rangle \right] $, at the
cost of the lost of unitarity $d\left\langle M\right. \left| M\right\rangle
/d\tau <0$ induced by the decay operator $j\widehat{\gamma }$. This
restriction of the Hilbert space to $\left[ \left| M^{0}\right\rangle
,\left| \overline{M}^{0}\right\rangle \right] $, allowed by the WW
approximation, leads to the effective LOY Hamiltonian. The LOY Hamiltonian
without CPV is the sum of the mass energy ($mc^{2}$), plus a
strangeness/bottomness $(S=\pm 1$ / $B^{\prime }=\pm 1$) coupling operator ($%
\widehat{\delta m}c^{2}$), plus the WW irreversible decay ($j\hbar \widehat{%
\gamma }$), according to the Schr\H{o}dinger equation 
\begin{equation}
j\hbar \frac{d\left| M\left( \tau \right) \right\rangle }{d\tau }%
=mc^{2}\left| M\left( \tau \right) \right\rangle -\left[ \widehat{\frac{%
\delta m}{2}}c^{2}+j\hbar \frac{\widehat{\gamma }}{2}\right] \cdot \left|
M\left( \tau \right) \right\rangle \text{.}  \label{a2}
\end{equation}
The coupling operator $\widehat{\delta m}$ and the decay operator $\widehat{%
\gamma }$ are given by 
\begin{eqnarray}
\widehat{\delta m} &=&\delta m\left[ \left| M^{0}\right\rangle \left\langle 
\overline{M}^{0}\right| +\left| \overline{M}^{0}\right\rangle \left\langle
M^{0}\right| \right] \text{,}  \label{a3} \\
\widehat{\gamma } &=&\Gamma \left[ \left| M^{0}\right\rangle \left\langle
M^{0}\right| +\left| \overline{M}^{0}\right\rangle \left\langle \overline{M}%
^{0}\right| \right]   \nonumber \\
&&-\delta \Gamma \left[ \left| M^{0}\right\rangle \left\langle \overline{M}%
^{0}\right| +\left| \overline{M}^{0}\right\rangle \left\langle M^{0}\right|
\right] \text{,}  \label{a4}
\end{eqnarray}
where $\delta m>0$ is the mass splitting between the heavy and light mass
eigenstates and $\Gamma >0$, $\delta \Gamma <0$ are respectively the average
and the splitting between the decay widths of the these eigenstates. These
mass eigenstates are: the long-lived $L$ and short-lived $S$ $\ $ states $%
\left( K_{S/L}\right) $ for $K^{0}/\overline{K}^{0}$, and the heavy $H$ and
light $L$ states $\left( B_{L/H}\right) $ for $B^{0}/\overline{B}^{0}$. We
take the convention $CP\left| M^{0}\right\rangle =\left| \overline{M}%
^{0}\right\rangle $. The CP eigenstates $\left| M_{1}\right\rangle $ and $%
\left| M_{2}\right\rangle $ are related to the flavor eigenstates by 
\begin{eqnarray}
\left| M_{1}\right\rangle  &=&\frac{\left| M^{0}\right\rangle }{\sqrt{2}}+%
\frac{\left| \overline{M}^{0}\right\rangle }{\sqrt{2}}=CP\left|
M_{1}\right\rangle \text{,}  \label{a5} \\
\left| M_{2}\right\rangle  &=&\frac{\left| M^{0}\right\rangle }{\sqrt{2}}-%
\frac{\left| \overline{M}^{0}\right\rangle }{\sqrt{2}}=-CP\left|
M_{2}\right\rangle \text{.}  \label{a6}
\end{eqnarray}
These CP eigenstates, $M_{1}$ and $M_{2}$, are also mass eigenstates, thus
the time evolution of the CP and mass eigenstates without CPV is given by 
\begin{eqnarray}
\left| M_{1}\left( \tau \right) \right\rangle  &=&\left| M_{1}\right\rangle
\exp -j\frac{c^{2}}{\hbar }\left[ m-\frac{\delta m}{2}-j\hbar \frac{\Gamma
-\delta \Gamma }{2c^{2}}\right] \tau \label{a7} \\
\left| M_{2}\left( \tau \right) \right\rangle  &=&\left| M_{2}\right\rangle
\exp -j\frac{c^{2}}{\hbar }\left[ m+\frac{\delta m}{2}-j\hbar \frac{\Gamma
+\delta \Gamma }{2c^{2}}\right] \tau \label{a8}
\end{eqnarray}
The above symmetric picture where CP commute with the Hamiltonian is no
longer valid when the experimental results of CPV are to be taken into
account.

\section{Mass eigenstates with types (\textit{i}) and (\textit{iii}) CPV}

When CPV comes into play, the Hamiltonian \ (\ref{a2}) is modified and\ the
mass eigenstates $K_{S/L}$ or $B_{L/H}$ are no longer the CP eigenstates $%
K_{1/2}$ or $B_{1/2}$\ (\ref{a5}, \ref{a6}).\ The mass eigenvalues are not
significantly changed by CPV.

For types ({\it i} and 
{\it iii}) CPV, experimental evidences
require to modify the Hamiltonian and the resulting mass eigenstates. Type ({\it ii})
direct CPV in the decay to one final state is also due to
earth's gravity, as will be demonstrated in section VI, but the associated $%
\varepsilon ^{\prime }$ parameter is not involved in the Hamiltonian
describing mixing. The parameters $\varepsilon $ and $\beta $ are introduced
in order to describe $K^{0}/\overline{K}^{0}$ type (\textit{i}) and $B^{0}/%
\overline{B}^{0}$ type (\textit{iii}) CPV effects.

For $K^{0}/\overline{K}^{0}$ type (\textit{i}) CPV, the indirect CPV effects
are described by the small complex parameter $\varepsilon $ and the mass
eigenstates $\left| K_{S/L}\right\rangle $ are related to the CP eigenstates 
$\left| K_{1/2}\right\rangle $ (\ref{a5}, \ref{a6}) by 
\begin{eqnarray}
\left| K_{S}\right\rangle  &=&\left| K_{1}\right\rangle +\varepsilon \left|
K_{2}\right\rangle \text{,}  \label{a9} \\
\left| K_{L}\right\rangle  &=&\left| K_{2}\right\rangle +\varepsilon \left|
K_{1}\right\rangle \text{.}  \label{a10}
\end{eqnarray}
As $\left| \varepsilon \right| =2.2\times 10^{-3}$ we have neglect $O\left[
10^{-6}\right] $ corrections associated with the normalization $\left\langle
K_{S/L}\right. \left| K_{S/L}\right\rangle $ $=$ $1$. The quantity $%
\left\langle K_{S/L}\right. \left| K_{L/S}\right\rangle =2\mathop{\rm Re}
\varepsilon $ is an observable.

For $B^{0}/\overline{B}^{0}$ type ({\it iii}) CPV, it is convenient to
consider mass eigenstates $\left| B_{L/H}\right\rangle $ related to CP
eigenstates $\left| B_{1/2}\right\rangle $ (\ref{a5},\ref{a6}) by 
\begin{eqnarray}
\left| B_{L}\right\rangle  &=&\cos \beta \left| B_{1}\right\rangle +j\sin
\beta \left| B_{2}\right\rangle \text{,}  \label{a11} \\
\left| B_{H}\right\rangle  &=&\cos \beta \left| B_{2}\right\rangle +j\sin
\beta \left| B_{1}\right\rangle \text{.}  \label{a12}
\end{eqnarray}

The overlap of mass eigenstates $\left\langle B_{L}\right. \left|
B_{S}\right\rangle =0$, thus there is no type (\textit{i}) CPV with this
parametrization and normalization is ensured as $\left\langle B_{L/H}\right.
\left| B_{L/H}\right\rangle $ $=$ $1$.

In the usual KM interpretation these parameters $\varepsilon $ and $\beta $
are related to combinations of CKM matrix elements where the KM phase is
adjusted to the measured CPV amplitude. Rather than adjusting a complex
phase, an other interpretation of the experiments is proposed below: we
simply take into account the impact of earth's gravity on the experiments
without the need to introduce a new parameters in a CP conserving CKM matrix
which is thus free of CPV far from any massive object in a physical
Lorentzian frame.

The final quantitative results predicted with this new mechanism leads to
the conclusion that CPV observed in the three canonical types of flavored
neutral mesons experiments (\textit{i, ii} and \textit{iii}) is (earth)
gravity induced, and not fundamental at the level of the CKM matrix elements.

\section{Strangeness and bottomness oscillations on earth}

The Schr\H{o}dinger equation Eq.(\ref{a2}) is pertinent far from any massive
object, but, on earth, we have to consider the very small Newtonian
potential energy $mG_{N}M_{\oplus }/R_{\oplus }$ $\sim $ $10^{-9}mc^{2}$. We
can restrict the description of this new coupling to the first term of the
Taylor expansion of \ $mG_{N}$ $M_{\oplus }/\left( R_{\oplus }+X+x\right) $
with respect to a vertical position $X$ $+$ $x$ where $x\ll X\ll R_{\oplus }$%
. The position$\ X$ is the vertical average position of the meson with
respect to the level $R_{\oplus }$. This is an external degree of freedom:
it can not enter in the $\tau $ dynamics (\ref{a2}) as $\tau $ is the meson
proper time. The vertical position $\ x\left( \tau \right) $ describes the
internal vertical fluctuations around this average $X$. This is an internal
degree of freedom: it must enter the proper time Hamiltonian (\ref{a2}).
Thus, we consider an additional energy term $mg\ \widehat{x}\left( \tau
\right) $ in (\ref{a2}) with $g$ = $G_{N}M_{\oplus }/R_{\oplus }^{2}$ = $9.8$
m/s$^{2}$,

\begin{eqnarray}
j\hbar \frac{d\left| M\left( \tau \right) \right\rangle }{d\tau }
&=&mc^{2}\left| M\right\rangle -\widehat{\frac{\delta m}{2}}c^{2}\cdot
\left| M\right\rangle  \nonumber \\
&&-j\hbar \frac{\widehat{\gamma }}{2}\cdot \left| M\right\rangle +mg\ \ 
\widehat{x}\left( \tau \right) \cdot \left| M\right\rangle \text{.}
\label{a13}
\end{eqnarray}

We have just applied here the \textit{correspondence principle} between
classical and quantum mechanics and this requires a careful interpretation
and analysis of the vertical position operator $\ \widehat{x}\left(\tau
\right) $.

It is very important to note that, as $\tau $ is the meson proper time, the
position operator $\widehat{x}\left( \tau \right) $ in (\ref{a13}) must not
be associated with the vertical position of the meson with respect to a
reference vertical level\ in the laboratory.

As $\tau $ is the rest frame proper time of the meson, the motion described
by the operator $\widehat{x}\left( \tau \right) $ is associated with the
(unknown) internal quark vertical motion, inside the mesons, as a function
of the meson proper time $\tau $: the zitterbewegung motion inherent to all,
free or bound, spin 1/2 fermions \cite{11}.

The operator $\widehat{x}\left( \tau \right) $ in (\ref{a13}) describes the
fast fluctuating vertical motion of the quarks inside the meson with respect
to the meson average position defining the rest frame of the meson. This
rest frame has a proper time $\tau $ and its free fall does not affect 
(\ref{a13}) on the time scale of the experiment.

The separation between a slow average and fast fluctuations is based on a
three time scales ordering of the dynamics: a very slow time scale of the
meson free fall, which does not enter in the meson proper time dynamics Eq.(%
\ref{a13}), a slow time scale of the order of $\ \hbar /\delta mc^{2}$
associated with  flavor oscillations, and the fast fluctuating motion
associated with quarks zitterbewegung internal oscillations, with a
zitterbewegung time scale of the order of the Compton wavelength divided 
by the velocity of light \textit{\ }$\lambda _{C}/c$ $\sim
\hbar /mc^{2}$. This separation between: average position, fast
zitterbewegung fluctuations around this average, and mixing oscillations,
displays a strong ordering. The mixing and zitterbewegung time scales
entering in (\ref{a13}) are ordered according to $ \hbar /mc^{2}:\hbar /\delta mc^{2}
\sim 10^{-15}-10^{-13}$.

The mesons, $\left| \overline{M}^{0}\right\rangle $ and $\left| \overline{M}%
^{0}\right\rangle $, are stationary bound states ultimately described by
Dirac spinors associated with one light quark $q^{\prime }$ and one heavier
quark $q$: $\left| M^{0}\right\rangle \sim \left| q^{\prime }\overline{q}%
\right\rangle $ and $\left| \overline{M}^{0}\right\rangle $ $\sim $ $\left| 
\overline{q}^{\prime }q\right\rangle $. The Dirac spinors are combined into a 
\textit{singlet}
spin zero $\left|q^{\prime }\overline{q}\right\rangle $ state. 
In the restricted Hilbert space, $\left[ \left|
M^{0}\right\rangle ,\left| \overline{M}^{0}\right\rangle \right] $, the
internal vertical position operator $\widehat{x}\left( \tau \right) $ is
thus represented by the following four matrix elements $\left\langle
{}\right| \widehat{x}\left| {}\right\rangle $ of the stationary Dirac
spinors $\left| q^{\prime }\overline{q}\right\rangle $ describing
quark-antiquark bound states

\begin{eqnarray}
\widehat{x\left( \tau \right) } &=&\left\langle q^{\prime }\overline{q}%
\right| \widehat{x}\left( \tau \right) \left| q^{\prime }\overline{q}%
\right\rangle \left| M^{0}\right\rangle \left\langle M^{0}\right|  \nonumber
\\
&&+\left\langle \overline{q}^{\prime }q\right| \widehat{x}\left( \tau
\right) \left| \overline{q}^{\prime }q\right\rangle \left| \overline{M}%
^{0}\right\rangle \left\langle \overline{M}^{0}\right|  \nonumber \\
&&+\left\langle \overline{q}^{\prime }q\right| \widehat{x}\left( \tau
\right) \left| q^{\prime }\overline{q}\right\rangle \left| \overline{M}%
^{0}\right\rangle \left\langle M^{0}\right|  \nonumber \\
&&+\left\langle q^{\prime }\overline{q}\right| \widehat{x}\left( \tau
\right) \left| \overline{q}^{\prime }q\right\rangle \left|
M^{0}\right\rangle \left\langle \overline{M}^{0}\right| \text{.}  \label{a14}
\end{eqnarray}

The internal vertical position operator $\widehat{\mathbf{x}}\left( \tau
\right) $ fulfils Heisenberg's equation $j\hbar d\widehat{\mathbf{x}}/d\tau $
= $\widehat{\mathbf{x}}\cdot \widehat{H}$ - $\widehat{H}\cdot \widehat{%
\mathbf{x}}$ where $\widehat{H}$ is the Dirac Hamiltonian describing quarks
confinement inside the meson. The values of the internal vertical position
matrix elements $\left\langle {}\right| \widehat{x}\left( \tau \right)
\left| {}\right\rangle $ depend of the model of their confinement inside the
meson. 

The typical size of $\sqrt{\left\langle x^{2}\right\rangle }$ is the Compton
wavelength of the meson $\lambda _{C}$.

The instantaneous velocity operator $d\widehat{\mathbf{x}}/d\tau $ of spin $%
1/2$ particles/antiparticles pairs, either free or bound, is well known to
display the so called zitterbewegung (nonintuitive) behavior: a quiver
motion on a length scale given by the Compton wavelength $\lambda _{C}$ and
an instantaneous velocity equal to the velocity light $c$ \cite{11}.

It is important to note that the values of the instantaneous velocity matrix
elements $\left\langle {}\right| d\widehat{\mathbf{x}}/d\tau \left|
{}\right\rangle $ are independent of the charge and mass of the fermions as
well as of the shape and strength of the effective confinement potential
involved in the Dirac Hamiltonian $\widehat{H}\left( \mathbf{x},\mathbf{p}%
\right) $. This zitterbewegung universality is a consequence of Heisenberg's
equation 
\begin{equation}
j\hbar \frac{d\mathbf{x}}{d\tau }=\left[ \mathbf{x},\widehat{H}\left( 
\mathbf{x},\mathbf{p}\right) \right] =j\hbar c{\boldsymbol \alpha }\text{.}
\label{a15}
\end{equation}
We have introduced the usual $4\times 4$ alpha matrices: ${\boldsymbol \alpha }$ $=$ $%
\left( \alpha _{x},\alpha _{y},\alpha _{z}\right) $ \cite{11} which can be
expressed in terms of the $2\times 2$ Pauli matrices ${\boldsymbol \sigma }$ $=$ $\left(
\sigma _{x},\sigma _{y},\sigma _{z}\right) $. Equation (\ref{a15}) imply
that the values of the internal fluctuating velocity matrix elements are
equal to $c$ or $0$. In two previous studies, Ref. \cite{9} and \cite{12},
we have found, with two different methods, that the normalized spinor bound
states in a confining potential describing the mean field strong
interactions ultimately ensuring confinement in the meson, fulfils $%
\left\langle \overline{q}\right| \alpha _{x}\left| \overline{q}\right\rangle 
$ = $\left\langle q\right| \alpha _{x}\left| q\right\rangle $ = $0$ and $%
\left\langle q\right| \alpha _{x}\left| \overline{q}\right\rangle $ = $%
\left\langle \overline{q}\right| \alpha _{x}\left| q\right\rangle $ = $1$.
To adapt the LOY model to earth's gravity, and to describe a meson and its
antimeson, we have to consider that antiparticles are just particles
propagating backward in time according to the Feynman picture. Following
Feynman's rule, to discriminate $M^{0}$ from $\overline{M}^{0}$ in the LOY
model on earth, the internal, vertical, fluctuating, velocity operator $d%
\widehat{x}/d\tau $,  expressed on the $\left[ \left| M^{0}\right\rangle
,\left| \overline{M}^{0}\right\rangle \right] $ basis, is thus given by 
\begin{equation}
\frac{d\widehat{x}}{d\tau }=c\left| M^{0}\right\rangle \left\langle 
\overline{M}^{0}\right| -c\left| \overline{M}^{0}\right\rangle \left\langle
M^{0}\right| \text{.}  \label{a16}
\end{equation}

This operator describes the instantaneous velocity (fast time scale) of the
quarks, inside the meson with respect to the meson rest frame.

Beside the time ordering between an average and an instantaneous dynamics,
the inclusion of the gravity term in (\ref{a2}), to give (\ref{a13}),
introduces an energy ordering. The Compton wavelength of the meson $\lambda
_{C}$ provides an approximate maximum size of the $q/\overline{q}$ quark
matrix elements $\left| \left\langle {}\right| \widehat{x}\left|
{}\right\rangle \right| $ in (\ref{a14}) as quarks are bound states inside
the volume of a meson. The very small numerical value of the energy $%
mg\lambda _{C}=\hbar g/c\sim 10^{-23}$ eV in front of $\delta mc^{2}\sim
10^{-6}-10^{-4}$ eV leads to the occurrence of a very strong ordering
fulfilled by the four matrix elements in (\ref{a14}), $mg\left| \left\langle
{}\right| \widehat{x}\left| {}\right\rangle \right| $ $\sim mg\lambda _{C}$ $%
\ll \delta mc^{2}$, in front of the other LOY matrix elements. Note that,
beside this energy ordering, the frequency ordering between these two terms
is reversed: $\delta mc^{2}/\hbar \ll mc^{2}/\hbar $.

The very strong energy ordering identified here allows to set up a
perturbative expansion of \ (\ref{a13}) with respect to the small expansion
parameter $\hbar g/\delta mc^{3}$ $\sim 10^{-19}-10^{-17}$. We define $%
\left| N\left( \tau \right) \right\rangle $ and $\left| n\left( \tau \right)
\right\rangle $ such that the meson dynamics is described by $\left| N\left(
\tau \right) \right\rangle +\left| n\left( \tau \right) \right\rangle $

\begin{eqnarray}
\left| M\left( \tau \right) \right\rangle  &=&\left| N\left( \tau \right)
\right\rangle \exp -j\frac{mc^{2}\tau }{\hbar }  \nonumber \\
&&+\left| n\left( \tau \right) \right\rangle \exp -j\frac{mc^{2}\tau }{\hbar 
}\text{.}  \label{a17}
\end{eqnarray}
The states $\left| N\left( \tau \right) \right\rangle $ and $\left| n\left(
\tau \right) \right\rangle $ are ordered according to: $\ \left| N\text{ }%
\right\rangle \sim $ $O\left( \hbar g/\delta mc^{3}\right) ^{0}$, $\left|
n\right\rangle $ $\sim O\left( \hbar g/\delta mc^{3}\right) ^{1}$and the
first neglected term is $O\left( \hbar g/\delta mc^{3}\right) ^{2}$ $\sim $\ 
$10^{-38}-10^{-34}$. With this expansion scheme (\ref{a17}), Schr\"{o}%
dinger's equation (\ref{a13}) becomes 
\begin{eqnarray}
j\hbar \frac{d\left| N\right\rangle }{d\tau } &=&-\frac{1}{2}\left( \widehat{%
\delta m}c^{2}+j\hbar \widehat{\gamma }\right) \cdot \left| N\right\rangle 
\text{,}  \label{a18} \\
j\hbar \frac{d\left| n\right\rangle }{d\tau } &=&-\frac{1}{2}\left( \widehat{%
\delta m}c^{2}+j\hbar \widehat{\gamma }\right) \cdot \left| n\right\rangle
+mg\ \widehat{x}\cdot \left| N\right\rangle \text{.}  \label{a19}
\end{eqnarray}
To identify the dominant secular contribution of earth's gravity we
introduce the inverse of the operator $\widehat{\delta m}c^{2}+j\hbar 
\widehat{\gamma }$ and then use this operator and (\ref{a18}) to rewrite (%
\ref{a19}) 
\begin{eqnarray}
j\hbar \frac{d\left| n\right\rangle }{d\tau } &=&-\frac{1}{2}\left( \widehat{%
\delta m}c^{2}+j\hbar \widehat{\gamma }\right) \cdot \left| n\right\rangle  
\nonumber \\
&&+2jmg\hbar \frac{d\widehat{x}}{d\tau }\ \cdot \left( \widehat{\delta m}%
c^{2}+j\hbar \widehat{\gamma }\right) ^{-1}\cdot \left| N\right\rangle  
\nonumber \\
&&-2jmg\hbar \frac{d}{d\tau }\left[ \ \widehat{x}\cdot \left( \widehat{%
\delta m}c^{2}+j\hbar \widehat{\gamma }\right) ^{-1}\cdot \left|
N\right\rangle \right] \text{.}  \label{a20}
\end{eqnarray}
The strong time ordering between strangeness (or bottomness) oscillations $%
\left( \hbar /\delta mc^{2}\right) $ and zitterbewegung oscillations $\left(
\sim \hbar /mc^{2}\right) $ can be used to simplify (\ref{a20}). We are
interested by the strangeness or bottomness dynamics taking place on the 
\textit{slow} time scale $\hbar /\delta mc^{2}$, thus we introduce $2\theta $
the period of the (unknown) \textit{fast} periodic functions $\left\langle
{}\right| \widehat{x}\left( \tau \right) \left| {}\right\rangle $ associated
with the zitterbewegung oscillations. This time $2\theta $ is such that $%
\hbar /mc^{2}\sim \theta \ll $ $\hbar /\delta mc^{2}$. We apply the
averaging operator $\widehat{A}_{\theta }\equiv $\ $\int_{\tau-\theta }^{\tau
+\theta }dt/2\theta $ on both side of (\ref{a20}) to average the high
frequency $\left( mc^{2}/\hbar \right) $ components. For any low frequency $%
\left( \delta mc^{2}/\hbar \right) $ function $f\left( t\right) $: $\widehat{%
A}_{\theta }\cdot f\left( t\right) =f\left( \tau \right) $ and $\widehat{A}%
_{\theta }\cdot df/dt=$ $df/d\tau $ and for any high frequency function $%
g\left( t\right) $: $\widehat{A}_{\theta }\cdot dg/dt=0$ . This usual
averaging methods is just \textit{Bogolioubov-Krilov-Mitropolski} method
when applied on the dynamical equations, or \textit{Witham} method if we
average directly the Lagrangian associated with the evolution equations \cite
{13}.

The equations describing strangeness or bottomness oscillations of a neutral
meson $\left| N\left( \tau \right) \right\rangle +\left| n\left( \tau
\right) \right\rangle $ on earth are given by 
\begin{eqnarray}
j\hbar \frac{d\left| N\right\rangle }{d\tau } &=&-\frac{1}{2}\left( \widehat{%
\delta m}c^{2}+j\hbar \widehat{\gamma }\right) \cdot \left| N\right\rangle 
\text{,}  \label{a21} \\
j\hbar \frac{d\left| n\right\rangle }{d\tau } &=&-\frac{1}{2}\left( \widehat{%
\delta m}c^{2}+j\hbar \widehat{\gamma }\right) \cdot \left| n\right\rangle +j%
\widehat{G}\cdot \left| N\right\rangle \text{.}  \label{a22}
\end{eqnarray}
The \textit{gravity/zitterbewegung} operator $\widehat{G}$, capturing the
secular interplay between zitterbewegung oscillations and bottomness or
strangeness oscillations, is defined as 
\begin{equation}
\widehat{G}=2mg\hbar \ \left( \frac{d\widehat{x}}{d\tau }\right) \cdot
\left( \widehat{\delta m}c^{2}+j\hbar \widehat{\gamma }\right) ^{-1}\text{.}
\label{a23}
\end{equation}

Flavored neutral mesons pairs $K^{0}/\overline{K}^{0}$ and $B^{0}/\overline{B%
}^{0}$ display different $m$, $\delta m$, $\Gamma $ and $\delta \Gamma $ and
the impact of earth gravity on their oscillation behavior is to be analyzed
specifically. In the following we keep the notation of \ Eqs.\ (\ref{a21}, 
\ref{a22}) and (\ref{a23}) with an additional index $K$ or $B$ for these
specific studies.

\section{Gravity induced type (\textit{i}) CPV in the mixing of $K^{0}/%
\overline{K}^{0}$}

The ordering associated with the specific case of a $K^{0}/\overline{K}^{0}$
pair is given by: $\delta m_{K}/m_{K}\sim 10^{-15}$ and the lifetime of the $%
K_{S}$ is $570$ times shorter than the lifetime of $\ K_{L}$. The first step
to interpret $K^{0}/\overline{K}^{0}$ experiments is to consider a unitary
evolution and to neglect the finite lifetime of both particles, described by
$j\hbar \widehat{\gamma }$, in Eqs. (\ref{a21}, \ref{a22}) and (\ref{a23}).
Then, as the lifetime of $K_{L}$ is 570 times longer than the lifetime of $%
K_{S}$, we set up a steady state balance between the fast decay of the small 
$K_{1}$ component of a $K_{L}$, produced initially without $K_{1}$, and its
gravity induced regeneration from this $K_{L}$. Considering first a unitary
evolution $\widehat{\gamma }=\widehat{0}$ $\left( \delta \Gamma _{K}=\Gamma
_{K}=0\right) $, we have to solve 
\begin{eqnarray}
j\hbar \frac{d\left| N_{K}\right\rangle }{d\tau } &=&-\frac{1}{2}\widehat{%
\delta m_{K}}c^{2}\cdot \left| N_{K}\right\rangle \text{,}  \label{a24} \\
j\hbar \frac{d\left| n_{K}\right\rangle }{d\tau } &=&-\frac{1}{2}\widehat{%
\delta m_{K}}c^{2}\cdot \left| n_{K}\right\rangle +j\widehat{G}_{K}\cdot
\left| N_{K}\right\rangle \text{.}  \label{a25}
\end{eqnarray}
The operator $\widehat{\delta m_{K}}c^{2}$ is given by (\ref{a3}), the
operator $d\widehat{x}/d\tau $ by (\ref{a16}), $\widehat{G}_{K}$ by (\ref
{a23}) and $\widehat{\gamma }=\widehat{0}$. The action of $\widehat{G}_{K}$
on the CP eigenstates $\left| K_{1}\right\rangle $ and $\left|
K_{2}\right\rangle $ defined in (\ref{a5}, \ref{a6}) is 
\begin{eqnarray}
\widehat{G}_{K}\left| K_{2}\right\rangle  &=&\kappa \delta m_{K}c^{2}\left|
K_{1}\right\rangle \text{,}  \label{a26} \\
\widehat{G}_{K}\left| K_{1}\right\rangle  &=&\kappa \delta m_{K}c^{2}\left|
K_{2}\right\rangle \text{,}  \label{a27}
\end{eqnarray}
where we have defined the small parameter $\kappa $ 
\begin{equation}
\kappa =2m_{K}g\hbar /\delta m_{K}^{2}c^{3}=1.7\times 10^{-3}\text{.}
\label{a28}
\end{equation}
This small parameter has been identified and discussed by Fishbach, forty
five years ago, as the undimensional combination matching approximately the
experimental value of $\mathop{\rm Re}\varepsilon $ \cite{14, 15}.

If we consider the following CP eigenstate 
\begin{equation}
\left| N_{K_{2}}\left( \tau \right) \right\rangle =\left| K_{2}\right\rangle
\exp -j\delta m_{K}c^{2}\tau /2\hbar \text{,}  \label{a29}
\end{equation}
which is the $\left( m_{K}+\delta m_{K}\right) $ mass eigenstate without CPV
(\ref{a6},\ref{a8}), it fulfils Eq. (\ref{a24}) and the associated solution
of Eq. (\ref{a25}) is 
\begin{equation}
\left| n_{K_{2}}\left( \tau \right) \right\rangle =j\kappa \left|
K_{1}\right\rangle \exp -j\delta m_{K}c^{2}\tau /2\hbar \text{.}  \label{a30}
\end{equation}
Thus, on earth, the mass eigenstates $\left| K_{2}^{\oplus }\right\rangle $
is not the CP eigenstates $\left| K_{2}\right\rangle $ (\ref{a6}), but the
sum of the previous solutions (\ref{a29}, \ref{a30}) 
\begin{equation}
\left| K_{2}^{\oplus }\right\rangle =\left| K_{2}\right\rangle +j\kappa
\left| K_{1}\right\rangle \text{.}  \label{a31}
\end{equation}
We neglect the small correction $O[10^{-6}]$ needed for normalization and
consider $\left\langle K_{2}^{\oplus }\right. \left| K_{2}^{\oplus
}\right\rangle $ = $1$. A similar result is obtained for the other $\left(
m_{K}-\delta m_{K}\right) $ mass eigenstate without CPV (\ref{a5},\ref{a7})
by taking 
\begin{equation}
\left| N_{K_{1}}\left( \tau \right) \right\rangle =\left| K_{1}\right\rangle
\exp j\delta m_{K}c^{2}\tau /2\hbar   \label{a32}
\end{equation}
as a source term on the right hand side of Eq. (\ref{a25}). This leads to a
gravity induced correction 
\begin{equation}
\left| n_{K_{1}}\left( \tau \right) \right\rangle =-j\kappa \left|
K_{2}\right\rangle \exp j\delta m_{K}c^{2}\tau /2\hbar \text{.}  \label{a33}
\end{equation}
\ The other mass eigenstates on earth is not the CP eigenstates $\left|
K_{1}\right\rangle $ (\ref{a5}), but the sum of the previous solutions (\ref
{a32}, \ref{a33}) 
\begin{equation}
\left| K_{1}^{\oplus }\right\rangle =\left| K_{1}\right\rangle -j\kappa
\left| K_{2}\right\rangle \text{.}  \label{a34}
\end{equation}
At the fundamental level of a unitary evolution, without decays, the impact
of earth's gravity appears as a CPT violation, with T conservation, \
because the indirect violation parameter $\left\langle K_{1}^{\oplus
}\right. \left| K_{2}^{\oplus }\right\rangle $ = $2j\kappa $ is imaginary 
\cite{4}, rather than a CP and T violation with CPT conservation requiring a
non zero real value \cite{4}.

We must now take into account the $K_{1}$ fast decay. This decay will change
the picture, qualitatively: an apparent CP and T violation, with CPT
conservation, is measured experimentally rather than a CPT one, and
quantitatively: with the right prediction of $\mathop{\rm Re}\varepsilon _{\text{ }%
}$which is slightly smaller than $\kappa $.

The previous results, Eqs. (\ref{a31}, \ref{a34}), allow to calculate the%
\textit{\ gravity induced transition amplitude }$\Omega _{2\rightarrow 1}$
describing the transition amplitude per unit time from $\left| K_{2}^{\oplus
}\right\rangle $ $\exp -j\delta m_{K}c^{2}\tau /2\hbar $ \ to $\left|
K_{1}^{\oplus }\right\rangle $ $\exp j\delta m_{K}c^{2}\tau /2\hbar $,

\begin{equation}
\Omega _{2\rightarrow 1}=\left\langle \frac{dK_{2}^{\oplus }}{d\tau }\left|
K_{1}^{\oplus }\left( \tau \right) \right. \right\rangle =\kappa \frac{%
\delta m_{K}c^{2}}{\hbar }\exp j\frac{\delta m_{K}c^{2}}{\hbar }\tau \text{.}
\label{a35}
\end{equation}
This can be viewed as a gravity induced \textit{oscillating regeneration}
competing with the short-lived kaon irreversible decay to the set of final
states $\left\{ \left| f\right\rangle \right\} $. This decay takes place at
a rate $\Gamma _{1\rightarrow f}/2=\left( \Gamma _{K}-\delta \Gamma
_{K}\right) /2$ $\sim $ $\Sigma _{f}\left| \left\langle f\right| \mathcal{T}%
\left| K_{1}\right\rangle \right| ^{2}$. Note that $\left| \Omega
_{2\rightarrow 1}\right| $ $\sim $ $O\left[ 10^{-3}\Gamma _{1\rightarrow
f}\right] $ so, starting from a pure $O\left[ 1\right] K_{2}$ population, an 
$O\left[ 10^{-3}\right] K_{1}$ steady state satellite will be observed.

We consider now a typical experiment dedicated to indirect CPV.
Experimentally $K_{1}$ and $K_{2}$ are first produced together in equal
amounts. Then, after few $1/$ $\Gamma _{1\rightarrow f}$ \ decay times, the
initial content of $\left| K_{1}\right\rangle $ disappears and a pure $%
\left| K_{2}\right\rangle $ state is expected. In fact, the state $\left|
K_{L\exp }\left( \tau \right) \right\rangle $ observed in such an\
experiment is not a pure $\left| K_{2}\right\rangle $ state. This $\left|
K_{L\exp }\left( \tau \right) \right\rangle $ state is a linear
superposition of $\left| K_{2}\right\rangle $, plus a small amount of $%
\left| K_{1}\right\rangle $, 
\begin{equation}
\left| K_{L\exp }\left( \tau \right) \right\rangle =a_{2}\left( \tau \right)
\left| K_{2}\right\rangle +a_{1}\left( \tau \right) \left|
K_{1}\right\rangle \text{,}  \label{a36}
\end{equation}
resulting from the balance between gravity induced regeneration (\ref{a35})
and irreversible decay. We assume that the $\left| K_{2}\right\rangle $
component is stable and that the depletion of its amplitude associated with
the \textit{gravitational regeneration} of $\left| K_{1}\right\rangle $ is
negligible so that $\left| a_{2}\right| $ = $1$ and 
\begin{equation}
a_{2}\left( \tau \right) =\exp -j\delta m_{K}c^{2}\tau /2\hbar \text{.}
\label{a37}
\end{equation}
The amplitude $a_{1}$ of $\left| K_{1}\right\rangle $ in (\ref{a36}) is
given by the steady-state balance between a decay at the (amplitude) rate $%
\Gamma _{1\rightarrow f}/2$ on the one hand, and a (gravity induced)
transition/regeneration $\Omega _{2\rightarrow 1}$ (\ref{a35}) from $\left|
K_{2}\right\rangle $ on the other hand 
\begin{equation}
a_{2}\left( \tau \right) \Omega _{2\rightarrow 1}=a_{1}\left( \tau \right) 
\frac{\Gamma _{1\rightarrow f}}{2}\text{.}  \label{a38}
\end{equation}
The solution is this equation is 
\begin{equation}
a_{1}\left( \tau \right) =\frac{\delta m_{K}c^{2}}{\hbar \Gamma _{1}/2}%
\kappa \exp j\delta m_{K}c^{2}\tau /2\hbar \text{,}  \label{a39}
\end{equation}
where we have dropped $\rightarrow f$ in $\Gamma _{1}$ to
simplify the notations. This short-lived $\left| K_{1}\right\rangle $
component is observed through its two pion decay \cite{1}. Thus the observed
long-lived mass eigenstate $\left| K_{L\exp }\right\rangle $, obtained after
few $1/$ $\Gamma _{1}$\ decay times away from a neutral kaons source, must
be represented by 
\begin{equation}
\left| K_{L\exp }\right\rangle =\left| K_{2}\right\rangle +\frac{\delta
m_{K}c^{2}}{\hbar \Gamma _{1}/2}\kappa \left| K_{1}\right\rangle \text{.}
\label{a40}
\end{equation}
This is the usual CPV parametrization of the kaon state Eq. (\ref{a10}). The
value of the indirect, gravity induced, CPV parameter, 
\begin{equation}
\mathop{\rm Re}\varepsilon _{\exp }=\frac{\delta m_{K}c^{2}}{\hbar \Gamma _{1}/2}%
\kappa =1.66\times 10^{-3}\text{,}  \label{a41}
\end{equation}
is thus in agreement with the experimental value, reported by Gershon and
Nir,\ page 290 of Ref. \cite{5} 
\begin{equation}
\mathop{\rm Re}\varepsilon =\left( 1.66\pm 0.02\right) \times 10^{-3}\text{.}
\end{equation}

We have taken into account here the finite lifetime of the short-lived kaon,
to complete this analysis we have also to take into account the decay of the
other mass eigenstate, and this will reveal the phase of $\varepsilon $.
Considering $\Gamma _{S}$ = $\Gamma _{K}-\delta \Gamma _{K}$ for $K_{1}$,
and $\Gamma _{L}$ = $\Gamma _{K}+\delta \Gamma _{K}$ for $K_{2}$ $(\delta
\Gamma _{K}<0$), beside the usual definition of decay rates $\Gamma _{S/L}$
= $\Sigma _{f}\left| \left\langle f\right| \mathcal{T}\left|
K_{S/L}\right\rangle \right| ^{2}$ in terms of transition amplitudes, Bell
and Steinberger \cite{16} have demonstrated a general relation based on
global unitarity starting from the evaluation of $\ d\left\langle M\right.
\left| M\right\rangle /d\tau $ at $\tau =0$. Using the fact that, for $K_{S}$%
, the sum over the final states $\Sigma _{f}$ \ is dominated (99.9\%) by $%
K_{S}\rightarrow $ $2\pi $ decays, more precisely by the $K_{S}$ $%
\rightarrow \left| I_{0}\right\rangle $ decays (95\%) to the isospin-zero
combination of $\left| \pi ^{+}\pi ^{-}\right\rangle $ and $\left| \pi
^{0}\pi ^{0}\right\rangle $, the Bell-Steinberger's unitarity relations \cite
{16} can be written:

\begin{equation}
j\frac{\delta mc^{2}}{\hbar }+\frac{\Gamma _{S}}{2}=\frac{\left\langle
I_{0}\right| \mathcal{T}\left| K_{L}\right\rangle \left\langle I_{0}\right| 
\mathcal{T}\left| K_{S}\right\rangle ^{*}}{\left\langle K_{S}\right. \left|
K_{L}\right\rangle }\text{.}  \label{a42}
\end{equation}
The restriction of the sum $\sum_{f}\left| f\right\rangle $ to $\left|
I_{0}\right\rangle $ reduces the $K_{S}$ width to $\Gamma _{S}$ = $%
\left\langle I_{0}\right| \mathcal{T}\left| K_{S}\right\rangle \left\langle
I_{0}\right| \mathcal{T}\left| K_{S}\right\rangle ^{*}$ so that 
\begin{equation}
\frac{\left\langle I_{0}\right| \mathcal{T}\left| K_{L}\right\rangle }{%
\left\langle I_{0}\right| \mathcal{T}\left| K_{S}\right\rangle }=\frac{%
\left\langle I_{0}\right| \mathcal{T}\left| K_{L}\right\rangle \left\langle
I_{0}\right| \mathcal{T}\left| K_{S}\right\rangle ^{*}}{\Gamma _{S}}\text{.}
\label{a43}
\end{equation}
This expression is then substituted in (\ref{a42}) to obtain the final
expression 
\begin{equation}
\frac{\left\langle I_{0}\right| \mathcal{T}\left| K_{L}\right\rangle }{%
\left\langle I_{0}\right| \mathcal{T}\left| K_{S}\right\rangle }=\frac{%
\left\langle K_{S}\right. \left| K_{L}\right\rangle }{2}\left( 1+j\frac{%
2\delta mc^{2}}{\hbar \Gamma _{S}}\right) \text{.}  \label{a44}
\end{equation}
The left hand side of \ Eq. (\ref{a44}) is just the definition of the
complex parameter $\varepsilon $ and $\left\langle K_{S}\right. \left|
K_{L}\right\rangle /2=\mathop{\rm Re}\varepsilon $, thus the argument of the CPV
complex parameter $\varepsilon $ is given by 
\begin{equation}
\arg \varepsilon =\arctan \left( 2\delta m_{K}c^{2}/\hbar \Gamma _{S}\right)
=43.4^{\circ }\text{,}  \label{a45}
\end{equation}
in agreement with the experimental results \cite {5}. This last relation (%
\ref{a45}) complete (\ref{a41}) and confirms that gravity induced CPV
provides a global pertinent framework to interpret $K^{0}/\overline{K}^{0}$
indirect CPV experiments.

\section{Gravity induced type (\textit{ii}) CPV in the decay of $K^{0}/%
\overline{K}^{0}$}

To interpret the measurements of the direct violation parameter $\varepsilon
^{\prime }$ we consider the $2\pi ^{0}$ decays of $K_{L}$ and $K_{S}$.

The definition of $\varepsilon ^{\prime }$ as a function of the amplitude
ratio $\eta _{00}$ is given by 
\begin{equation}
\eta _{00}=\frac{\left\langle \pi ^{0}\pi ^{0}\right| \mathcal{T}\left|
K_{L}\right\rangle }{\left\langle \pi ^{0}\pi ^{0}\right| \mathcal{T}\left|
K_{S}\right\rangle }\equiv \varepsilon -2\varepsilon ^{\prime }\text{,}
\label{a46}
\end{equation}
where the direct violation in the decay in one final state $\varepsilon
^{\prime }$ is a correction to the indirect violation in the mixing $%
\varepsilon \gg \varepsilon ^{\prime }$.

The various bra and ket in a quantum model are defined up to an unobservable
phase. The arbitrary conventional phases inherent to quantum theoretical
models are to be eliminated to define phase-convention-independent
observables. This requirement is a consequence of the objectivity of a
physical description, for example, within the framework of relativistic
dynamics, all the physical observables should be expressed as combinations of
Lorentz invariants and here, within the framework of quantum dynamics, all
the physical observables are to be expressed as combinations of
phase-convention-independent quantities. The definition of $\eta _{00}$ is
invariant under rephasing of the pions state $\left\langle \pi ^{0}\pi
^{0}\right| $, but not with respect to the rephasing of the kaons mass
eigenstates $\left| K_{L/S}\right\rangle $. We can define an amplitude ratio 
$\eta _{00}$ which is a phase-convention-independent quantity through a
multiplication with the\textit{\ }factor $\varphi _{K}$ 
\begin{equation}
\varphi _{K}=\frac{\left\langle K^{0}\right. \left| K_{S}\right\rangle }{%
\left\langle K^{0}\right. \left| K_{L}\right\rangle }\text{,}  \label{a47}
\end{equation}
as the $L/S$ phase of $\eta _{00}$ is compensated by the $S/L$ phase of $%
\varphi _{K}$.

If we consider the mass eigenstates (\ref{a9}, \ref{a10}) used for the usual
description of CPV 
\begin{eqnarray}
\left| K_{S}\right\rangle &=&\frac{1+\varepsilon }{\sqrt{2}}\left|
K_{0}\right\rangle +\frac{1-\varepsilon }{\sqrt{2}}\left| \overline{K}%
^{0}\right\rangle \text{,}  \label{a48} \\
\left| K_{L}\right\rangle &=&\frac{1+\varepsilon }{\sqrt{2}}\left|
K_{0}\right\rangle -\frac{1-\varepsilon }{\sqrt{2}}\left| \overline{K}%
^{0}\right\rangle \text{.}  \label{a49}
\end{eqnarray}
and those obtained at the fundamental level of an unitary evolution (\ref
{a31}, \ref{a34}) with gravity induced CPV 
\begin{eqnarray}
\left| K_{1}^{\oplus }\right\rangle &=&\frac{1-j\kappa }{\sqrt{2}}\left|
K^{0}\right\rangle +\frac{1+j\kappa }{\sqrt{2}}\left| \overline{K}%
^{0}\right\rangle \text{,}  \label{a50} \\
\left| K_{2}^{\oplus }\right\rangle &=&\frac{1+j\kappa }{\sqrt{2}}\left|
K^{0}\right\rangle -\frac{1-j\kappa }{\sqrt{2}}\left| \overline{K}%
^{0}\right\rangle \text{,}  \label{a51}
\end{eqnarray}
we obtain two different expressions of the rephasing factor $\varphi _{K}$.

For the usual CPV parametrization (\ref{a48}, \ref{a49}) with CPT
conservation 
\begin{equation}
\varphi _{K}=\frac{\left\langle K^{0}\right. \left| K_{S}\right\rangle }{%
\left\langle K^{0}\right. \left| K_{L}\right\rangle }=1\text{.}  \label{a52}
\end{equation}

For gravity induced CPV, we replace $\left| K_{S}\right\rangle $ and $\left|
K_{L}\right\rangle $ by $\left| K_{1}^{\oplus }\right\rangle $ and $\left|
K_{2}^{\oplus }\right\rangle $ and use (\ref{a50}, \ref{a51}) to obtain 
\begin{equation}
\varphi _{K}^{\oplus }=\frac{\left\langle K^{0}\right. \left| K_{1}^{\oplus
}\right\rangle }{\left\langle K^{0}\right. \left| K_{2}^{\oplus
}\right\rangle }=1-\left\langle K_{1}^{\oplus }\right. \left| K_{2}^{\oplus
}\right\rangle \text{.}  \label{a 53}
\end{equation}
where $O\left[ 10^{-6}\right] $ and higher orders terms are neglected. We
assume that the amplitude of $K_{0}\rightarrow \pi ^{0}\pi ^{0}$ is equal to
the amplitude of $\ \overline{K}^{0}\rightarrow \pi ^{0}\pi ^{0}$ because
the interaction between a $\left( \pi ^{0}\pi ^{0}\right) $ state and a
neutral kaon state, $K_{0}$ or $\overline{K}^{0}$, can not differentiate the 
$K^{0}$ from the $\overline{K}^{0}$ (a final state phase can be absorbed by
a proper phase convention between the $K_{0}$ and $\overline{K}^{0}$). The
ratio of amplitude $\eta _{00}^{\oplus }$ associated with the mass
eigenstates (\ref{a50}, \ref{a51}) describing gravity induced CPV is 
\begin{equation}
\eta _{00}^{\oplus }=\frac{\left\langle \pi ^{0}\pi ^{0}\right| \mathcal{T}%
\left| K_{2}^{\oplus }\right\rangle }{\left\langle \pi ^{0}\pi ^{0}\right| 
\mathcal{T}\left| K_{1}^{\oplus }\right\rangle }=\frac{\left\langle
K_{1}^{\oplus }\right. \left| K_{2}^{\oplus }\right\rangle }{2}\text{.}
\label{a54}
\end{equation}
We conclude that the physical observable $\eta _{00}\varphi _{K}$ on earth
is given by 
\begin{equation}
\eta _{00}^{\oplus }\varphi _{K}^{\oplus }=\frac{\left\langle K_{1}^{\oplus
}\right. \left| K_{2}^{\oplus }\right\rangle }{2}\left[ 1-2\frac{%
\left\langle K_{1}^{\oplus }\right. \left| K_{2}^{\oplus }\right\rangle }{2}%
\right] \text{.}  \label{a55}
\end{equation}
When we neglect the finite lifetime of the kaons, we have demonstrated, 
in the previous section, that the gravity induced mixing 
$\widehat{G}_{K}$, between $\left| K_{1}\right\rangle $ and $\left|
K_{2}\right\rangle $, leads to an apparent CPT violation with $\left\langle
K_{1}^{\oplus }\right. \left| K_{2}^{\oplus }\right\rangle /2=j\kappa $. When decays are taken into
account, the finite lifetime of $K_{1}$ was shown to induce a rotation from the
imaginary value $j\kappa $ to the real observed value $\left( \delta
m_{K}c^{2}/\hbar \Gamma _{1}/2\right) \kappa =\left\langle K_{S}\right.
\left| K_{L}\right\rangle /2$ (\ref{a41}). Taking into account these
dissipative effects, the observed phase-convention-independent amplitude ratio $\eta
_{00\exp }^{\oplus }\varphi _{K\exp }^{\oplus }$ measured in dedicated
experiments on earth \cite {17, 18, 19} is given by the use of $%
\left\langle K_{S}\right. \left| K_{L}\right\rangle $ instead of $%
\left\langle K_{1}^{\oplus }\right. \left| K_{2}^{\oplus }\right\rangle $ 
in the phase-convention-independent observable (\ref{a55}) 
\begin{equation}
\eta _{00\exp }^{\oplus }\varphi _{K\exp }^{\oplus }=\frac{\left\langle
K_{S}\right. \left| K_{L}\right\rangle }{2}\left[ 1-2\frac{\left\langle
K_{S}\right. \left| K_{L}\right\rangle }{2}\right] \text{.}  \label{a56}
\end{equation}

The definition (\ref{a46}) $\eta _{00}\varphi _{K}=\varepsilon -2\varepsilon
^{\prime }$ leads to the conclusion that $\mathop{\rm Re}\varepsilon ^{\prime }/\varepsilon
=\mathop{\rm Re}\varepsilon $. The gravity induced direct CPV parameter 
\begin{equation}
\mathop{\rm Re}\left( \varepsilon ^{\prime }/\varepsilon \right) =\frac{\delta
m_{K}c^{2}}{\hbar \Gamma _{1}/2}\kappa =1.66\times 10^{-3}\text{,}
\label{a57}
\end{equation}
is in agreement with the experimental value, reported by Gershon and Nir, \
page 285 of  Ref. \cite{5} 
\begin{equation}
\mathop{\rm Re}\left( \varepsilon ^{\prime }/\varepsilon \right) =\left( 1.66\pm
0.23\right) \times 10^{-3}\text{.}
\end{equation}
The precise definition of \ phase-convention-independent quantities, in
order to clearly identify what is measured in an experiment, is also one of
the key to interpret the experimental observation of interferences between
mixing and decay in dedicated $B^{0}/\overline{B}^{0}$ experiments.

\section{Gravity induced type (\textit{iii}) CPV in the interference between
mixing and decay of $B^{0}/\overline{B}^{0}$}

The mass and width ordering associated with the $B^{0}/\overline{B}^{0}$
system is given by : $\delta m_{B}/m_{B}\sim 10^{-19}$ and $\delta
m_{B}/\Gamma _{B}\sim 0.7$. The lifetime of the CP eigenstate $B_{1}$ is
considered equal to the lifetime of the other CP eigenstate $B_{2}$ so that $%
\delta \Gamma _{B}=0$. The most pronounced CPV effects in the $B^{0}/%
\overline{B}^{0}$ system is displayed through interference experiments
dedicated to the study of the phase difference between the decay path $%
B_{0}\rightarrow f$ and the decay path $B_{0}\rightarrow \overline{B}%
^{0}\rightarrow f$ \ \cite{20, 21, 22}. To set up an interpretation of these
experiments we keep a finite lifetime $\Gamma _{B}^{-1}$ for both particles
and consider the diagonal decay operator 
\begin{equation}
\widehat{\gamma }_{B}=\Gamma _{B}\left[ \left| B^{0}\right\rangle
\left\langle B^{0}\right| +\left| \overline{B}^{0}\right\rangle \left\langle 
\overline{B}^{0}\right| \right] \text{,}  \label{a58}
\end{equation}
to describe the dissipative part of \ the bottomness dynamics. Thus, we have
to solve Eqs. (\ref{a21}, \ref{a22}) 
\begin{eqnarray}
j\hbar \frac{d\left| N_{B}\right\rangle }{d\tau } &=&-\frac{1}{2}\left( 
\widehat{\delta m_{B}}c^{2}+j\hbar \widehat{\gamma }_{B}\right) \cdot \left|
N_{B}\right\rangle \text{,}  \label{a59} \\
j\hbar \frac{d\left| n_{B}\right\rangle }{d\tau } &=&-\frac{1}{2}\left( 
\widehat{\delta m_{B}}c^{2}+j\hbar \widehat{\gamma }_{B}\right) \cdot \left|
n_{B}\right\rangle   \nonumber \\
&&+j\widehat{G}_{B}\cdot \left| N_{B}\right\rangle \text{.}  \label{a60}
\end{eqnarray}
The operator $\widehat{\delta m_{B}}c^{2}$ is given by (\ref{a3}), $\widehat{%
G}_{B}$ by (\ref{a23}), the operator $d\widehat{x}/d\tau $ by (\ref{a16}),
and $\widehat{\gamma} _{B}$ by (\ref{a58}). The action of $\widehat{G}_{B}$
on the CP eigenstates $\left| B_{1}\right\rangle $ and $\left|
B_{2}\right\rangle $ (\ref{a5}, \ref{a6}) is 
\begin{eqnarray}
j\widehat{G}_{B}\left| B_{2}\right\rangle  &=&-\delta m_{B}c^{2}\varkappa
\left( 1-j\chi \right) \left| B_{1}\right\rangle \text{,}  \label{a61} \\
j\widehat{G}_{B}\left| B_{1}\right\rangle  &=&\delta m_{B}c^{2}\varkappa
\left( 1+j\chi \right) \left| B_{2}\right\rangle \text{.}  \label{a62}
\end{eqnarray}
Where we define the real parameters $\chi $ and $\varkappa $ associated with
this gravity induced mixing of the $\left[ \left| B_{1}\right\rangle ,\left|
B_{2}\right\rangle \right] $ CP basis 
\begin{eqnarray}
\chi  &=&\delta m_{B}c^{2}/\hbar \Gamma _{B}=0.77\text{,}  \label{a63} \\
\varkappa  &=&2m_{B}g\hbar /\delta m_{B}^{2}c^{3}\left( \chi +\chi
^{-1}\right) \sim O\left[ 10^{-6}\right] \text{.}  \label{a64}
\end{eqnarray}
In order to solve Eq. (\ref{a60}) and to express the mass eigenstates on
earth, we consider the CP eigenstates 
\begin{equation}
\left| N_{B_{2}}\left( \tau \right) \right\rangle =\left| B_{2}\right\rangle
\exp -j\frac{\delta m_{B}c^{2}-j\hbar \Gamma _{B}}{2\hbar }\tau \text{,}
\label{a65}
\end{equation}
which is also the $\left( m_{B}+\delta m_{B}\right) $ mass eigenstate
without CPV, it fulfils (\ref{a59}) and the associated solution of \ (\ref
{a60}) is 
\begin{equation}
\left| n_{B_{2}}\left( \tau \right) \right\rangle =-\varkappa \left( 1-j\chi
\right) \left| B_{1}\right\rangle \exp -j\frac{\delta m_{B}c^{2}-j\hbar
\Gamma _{B}}{2\hbar }\tau \text{.}  \label{a66}
\end{equation}
Then we consider the other $\left( m_{B}-\delta m_{B}\right) $ CP eigenstate
 as a drive on the right hand side of Eq. (\ref{a60}) 
\begin{equation}
\left| N_{B_{1}}\left( \tau \right) \right\rangle =\left| B_{1}\right\rangle
\exp j\frac{\left( \delta m_{B}c^{2}+j\hbar \Gamma _{B}\right) }{2\hbar }%
\tau \text{.}  \label{a67}
\end{equation}
It fulfils Eq. (\ref {a59}) and the driven solution of \ Eq. (\ref{a60}) is 
\begin{equation}
\left| n_{B_{1}}\left( \tau \right) \right\rangle =-\varkappa \left( 1+j\chi
\right) \left| B_{2}\right\rangle \exp j\frac{\delta m_{B}c^{2}+j\hbar
\Gamma _{B}}{2\hbar }\tau \text{.}  \label{a68}
\end{equation}
Thus, on earth, the CP eigenstates $\left| B_{1}\right\rangle $ and $\left|
B_{2}\right\rangle $\ (\ref{a5}, \ref{a6}) are no longer the mass eigenstates 
$B_{L/H}^{\oplus }$ which are given by the sum $\left|
N_{B_{1/2}}\right\rangle +\left| n_{B_{1/2}}\right\rangle $ of (\ref{a65}, %
\ref{a67}) plus (\ref{a66}, \ref{a68}) 
\begin{eqnarray}
\left| B_{L}^{\oplus }\right\rangle  &=&\left| B_{1}\right\rangle -\varkappa
\left( 1+j\chi \right) \left| B_{2}\right\rangle \text{,}  \label{a69} \\
\left| B_{H}^{\oplus }\right\rangle  &=&\left| B_{2}\right\rangle -\varkappa
\left( 1-j\chi \right) \left| B_{1}\right\rangle \text{.}  \label{a70}
\end{eqnarray}
Using the bottomness basis $\left[ \left| B^{0}\right\rangle ,\left| 
\overline{B}^{0}\right\rangle \right] $ rather than the CP basis $\left[
\left| B_{1}\right\rangle ,\left| B_{2}\right\rangle \right] $ these mass
eigenstates become 

\begin{eqnarray}
\left| B_{L}^{\oplus }\right\rangle  &=&\frac{1-\varkappa \left( 1+j\chi
\right) }{\sqrt{2}}\left| B^{0}\right\rangle \nonumber \\
&&+\frac{1+\varkappa \left(1+j\chi \right) }{\sqrt{2}}\left| 
\overline{B}^{0}\right\rangle \text{,}
\label{a71}
\\
\left| B_{H}^{\oplus }\right\rangle  &=&\frac{1-\varkappa \left( 1-j\chi
\right) }{\sqrt{2}}\left| B^{0}\right\rangle \nonumber \\
&& -\frac{1+\varkappa \left(
1-j\chi \right) }{\sqrt{2}}\left| \overline{B}^{0}\right\rangle \text{.}
\label{a72}
\end{eqnarray}

The difference between these gravity induced mass eigenstates (\ref{a69}, 
\ref{a70}, \ref{a71}, \ref{a72}) and the usual type (\textit{iii})\ $B^{0}/%
\overline{B}^{0}$ parametrization (\ref{a11}, \ref{a12}),\ is that
gravity induced CPV requires two real number $\varkappa $ and $\chi $ to
express the eigenstates $\left| B_{L/H}^{\oplus }\right\rangle $ although
type (\textit{iii}) standard CPV parametrization (\ref{a11}, \ref{a12}) 
\begin{eqnarray}
\left| B_{L}\right\rangle &=&\frac{\exp +j\beta }{\sqrt{2}}\left|
B^{0}\right\rangle +\frac{\exp -j\beta }{\sqrt{2}}\left| \overline{B}%
^{0}\right\rangle \text{,}  \label{a73} \\
\left| B_{H}\right\rangle &=&\frac{\exp +j\beta }{\sqrt{2}}\left|
B^{0}\right\rangle -\frac{\exp -j\beta }{\sqrt{2}}\left| \overline{B}%
^{0}\right\rangle \text{,}  \label{a74}
\end{eqnarray}
is based on a single real parameter $\beta $ to interpret the experimental
results.

This difference is due to the assumption of CPT invariance associated with
the parametrization (\ref{a11}, \ref{a12}) and (\ref{a73}, \ref{a74}). There
are \textit{a priori} 4 complex numbers $\left\langle B^{0}\right. \left|
B_{L}\right\rangle $, $\left\langle \overline{B}^{0}\right. \left|
B_{H}\right\rangle $, $\left\langle B^{0}\right. \left| B_{H}\right\rangle $
and $\left\langle \overline{B}^{0}\right. \left| B_{L}\right\rangle $ needed
to expand a mass eigenstates $\left| B_{L/H}\right\rangle $ on a bottomness
basis $\left[ \left| B^{0}\right\rangle ,\left| \overline{B}%
^{0}\right\rangle \right] $, and this number is then reduced through
normalization and rephasing invariance.

When the decay into one final CP eigenstate $\left| f\right\rangle $\ is
considered in experiments, the observable is assumed to be the  combination
of amplitudes $\lambda _{f}$ defined as 
\begin{equation}
\lambda _{f}=\frac{\left\langle \overline{B}^{0}\right. \left|
B_{L}\right\rangle }{\left\langle B^{0}\right. \left| B_{L}\right\rangle }%
\frac{\left\langle f\right| \mathcal{T}\left| \overline{B}^{0}\right\rangle 
}{\left\langle f\right| \mathcal{T}\left| B^{0}\right\rangle }=\exp -2j\beta 
\frac{\left\langle f\right| \mathcal{T}\left| \overline{B}^{0}\right\rangle 
}{\left\langle f\right| \mathcal{T}\left| B^{0}\right\rangle }\text{,}
\label{a75}
\end{equation}
which is obviously phase-convention-independent. This parameter is observable 
through the measurement of $S_{f}$ and $C_{f}$ \ 
\begin{equation}
S_{f}=\frac{2\mathop{\rm Im}\lambda _{f}}{1+\left| \lambda _{f}\right| ^{2}}\text{%
, }C_{f}=\frac{1-\left| \lambda _{f}\right| ^{2}}{1+\left| \lambda
_{f}\right| ^{2}}\text{,}  \label{a76}
\end{equation}
which can be extracted from the data obtained from interferences between the
direct path $B_{0}\rightarrow f$ and the mixed path $B_{0}\rightarrow 
\overline{B}^{0}\rightarrow f$ .

This parameter $\lambda _{f}$ is meaningful to characterize type (\textit{iii%
}) CPV with the CPT invariant parametrization (\ref{a73}, \ref{a74}) because
it captures all the component of the expansion of the mass eigenstates on
the bottomness basis as 
\begin{equation}
\frac{\left\langle \overline{B}^{0}\right. \left| B_{L}\right\rangle }{%
\left\langle B^{0}\right. \left| B_{L}\right\rangle }=-\frac{\left\langle 
\overline{B}^{0}\right. \left| B_{H}\right\rangle }{\left\langle
B^{0}\right. \left| B_{H}\right\rangle }=\frac{\left\langle \overline{B}%
^{0}\right. \left| B_{L}\right\rangle }{\left\langle B^{0}\right. \left|
B_{H}\right\rangle }=-\frac{\left\langle \overline{B}^{0}\right. \left|
B_{H}\right\rangle }{\left\langle B^{0}\right. \left| B_{L}\right\rangle }%
\text{.}  \label{a77}
\end{equation}
These four amplitudes ratio are different if we consider the gravity induced
mass eigenstates (\ref{a71}, \ref{a72}) 
\begin{equation}
\frac{\left\langle \overline{B}^{0}\right. \left| B_{L}^{\oplus
}\right\rangle }{\left\langle B^{0}\right. \left| B_{L}^{\oplus
}\right\rangle }\neq -\frac{\left\langle \overline{B}^{0}\right. \left|
B_{H}^{\oplus }\right\rangle }{\left\langle B^{0}\right. \left|
B_{H}^{\oplus }\right\rangle }\neq \frac{\left\langle \overline{B}%
^{0}\right. \left| B_{L}^{\oplus }\right\rangle }{\left\langle B^{0}\right.
\left| B_{H}^{\oplus }\right\rangle }\neq -\frac{\left\langle \overline{B}%
^{0}\right. \left| B_{H}^{\oplus }\right\rangle }{\left\langle B^{0}\right.
\left| B_{L}^{\oplus }\right\rangle }\text{.}  \label{a78}
\end{equation}

Despite this difference between (\ref{a77}) and (\ref{a78}), the
experimental results analyzed within a CPT invariant framework (\ref{a77}),
can be understood and explained within the framework of gravity induced CPV (%
\ref{a78}). This situation is similar to the one encountered in section VI
devoted to the study of $\varepsilon ^{\prime }$: if CPT is assumed the
rephasing factors $\varphi =1$, and the interpretation of the experimental
measurements is based on the hypothesis of direct violation and imply a CPV
at the fundamental level of the CKM matrix. However, if earth's gravity
effects are taken into account $\varphi \neq 1$ and the very same
phase-convention-independent measured quantities agree with the experiments
without any additional assumptions. Earth's gravity was identified as the
sole source of $\varepsilon ^{\prime }$.

The analysis below will use two different approaches to interpret the
measurement of $\ \beta $, each providing the same final result. The two
issues addressed below are: first, the invariance under rephasing of the
mass eigenstates, when needed, to define an observable and second, the
invariance under rephasing of the flavor eigenstates, when needed, to define
an observable.

In order to accommodate the relation (\ref{a75}) with (\ref{a77}, \ref{a78}%
), we consider a $\lambda _{f}$ \ parameter constructed with the amplitude
ratio $\left\langle \overline{B}^{0}\right. \left| B_{L}\right\rangle
/\left\langle B^{0}\right. \left| B_{H}\right\rangle $ which is better
suited to characterize the dynamics of oscillating $B_{L/S}$ as it takes
into account all the eigenstates: the two flavor eigenstates and the two
mass eigenstates. However, this $\widetilde{\lambda }_{f}$ $\ $parameter
reflecting the $B_{L/S}$ content of the oscillating and propagating $B^{0}/%
\overline{B}^{0}$, 
\begin{equation}
\widetilde{\lambda }_{f}=\frac{\left\langle \overline{B}^{0}\right. \left|
B_{L}\right\rangle }{\left\langle B^{0}\right. \left| B_{H}\right\rangle }%
\frac{\left\langle f\right| \mathcal{T}\left| \overline{B}^{0}\right\rangle 
}{\left\langle f\right| \mathcal{T}\left| B^{0}\right\rangle }=\exp -2j\beta 
\frac{\left\langle f\right| \mathcal{T}\left| \overline{B}^{0}\right\rangle 
}{\left\langle f\right| \mathcal{T}\left| B^{0}\right\rangle }\text{,}
\label{a79}
\end{equation}
is not phase-convention-independent with respect to the mass eigenstates. 
The decay amplitude ratio $\left\langle f\right| \mathcal{T}\left| 
\overline{B}^{0}\right\rangle /\left\langle f\right| \mathcal{T}\left|
B^{0}\right\rangle $ provides invariance under rephasing of the flavor
eigenstates. Both flavor eigenstates $\overline{B}^{0}$ and $B^{0}$ and both
mass eigenstates $B_{L}^{\oplus }$ and $B_{H}^{\oplus }$ \ are involved in
real experiments, thus the amplitude ratio $\widetilde{\lambda }_{f}$
displaying these four states is more likely to provide the right
interpretation of the experimental measurement.

In order to set up a fully phase-convention-independent interpretation of
the experiments $\ $we introduce the symmetric rephasing factor 
\begin{equation}
\varphi _{B}=\sqrt{\frac{\left\langle B_{1}\right. \left| B_{H}\right\rangle 
}{\left\langle B_{1}\right. \left| B_{L}\right\rangle }\frac{\left\langle
B_{2}\right. \left| B_{H}\right\rangle }{\left\langle B_{2}\right. \left|
B_{L}\right\rangle }}=1\text{.}  \label{a80}
\end{equation}
We have used $B_{1/2}$ states because they are the CP eigenstates like $f$ \
which is also assumed to be a CP eigenstate. The amplitude ratio observed in
the experimental measurement are given by phase-convention-independent
product $\widetilde{\lambda }_{f}\varphi _{B}$ 
\begin{equation}
\widetilde{\lambda }_{f}\varphi _{B}=\frac{\left\langle \overline{B}%
^{0}\right. \left| B_{L}\right\rangle }{\left\langle B^{0}\right. \left|
B_{H}\right\rangle }\frac{\left\langle f\right| \mathcal{T}\left| \overline{B%
}^{0}\right\rangle }{\left\langle f\right| \mathcal{T}\left|
B^{0}\right\rangle }\varphi _{B}=\exp -2j\beta \frac{\left\langle f\right| 
\mathcal{T}\left| \overline{B}^{0}\right\rangle }{\left\langle f\right| 
\mathcal{T}\left| B^{0}\right\rangle }  \label{a81}
\end{equation}
which is equal to $\lambda _{f}$ (\ref{a75}).

When the same rephasing factor $\varphi _{B}^{\oplus }$ is calculated within
the framework of gravity induced CPV with (\ref{a69}, \ref{a70}) rather than
(\ref{a11}, \ref{a12}), this gives

\begin{equation}
\varphi _{B}^{\oplus }=\sqrt{\frac{\left\langle B_{1}\right. \left|
B_{H}^{\oplus }\right\rangle }{\left\langle B_{1}\right. \left|
B_{L}^{\oplus }\right\rangle }\frac{\left\langle B_{2}\right. \left|
B_{H}^{\oplus }\right\rangle }{\left\langle B_{2}\right. \left|
B_{L}^{\oplus }\right\rangle }}=\sqrt{\frac{1-j\chi }{1+j\chi }}\text{.}
\label{a82}
\end{equation}
The phase-convention-independent product, 
\begin{equation}
\widetilde{\lambda }_{f}^{\oplus }\varphi _{B}^{\oplus }=\frac{\left\langle 
\overline{B}^{0}\right. \left| B_{L}^{\oplus }\right\rangle }{\left\langle
B^{0}\right. \left| B_{H}^{\oplus }\right\rangle }\frac{\left\langle
f\right| \mathcal{T}\left| \overline{B}^{0}\right\rangle }{\left\langle
f\right| \mathcal{T}\left| B^{0}\right\rangle }\varphi _{B}^{\oplus }\text{,}
\label{a83}
\end{equation}
calculated with (\ref{a71}, \ref{a72}, \ref{a82}), becomes 
\begin{equation}
\widetilde{\lambda }_{f}^{\oplus }\varphi _{B}^{\oplus }=\exp \left(
-j\arctan \chi \right) \frac{\left\langle f\right| \mathcal{T}\left| 
\overline{B}^{0}\right\rangle }{\left\langle f\right| \mathcal{T}\left|
B^{0}\right\rangle }\left( 1+O\left[ 10^{-6}\right] \right) \text{.}
\label{a84}
\end{equation}
To compare the interpretations based on the usual CPT eigenstates $\left|
B_{L/H}\right\rangle $ (\ref{a81}) with the one based on gravity induced mass
eigenstates $\left| B_{L/H}^{\oplus }\right\rangle $ (\ref{a84}),
we must define $\beta $ such as $2\beta =$ $\arctan \left( 0.77\right) $. 

If $\arg \left( \left\langle f\right| \mathcal{T}\left| \overline{B}%
^{0}\right\rangle /\left\langle f\right| \mathcal{T}\left|
B^{0}\right\rangle \right) $ $=$ $0$ the experiments dedicated to $\overline{%
B}^{0}/B_{0}\rightarrow f$ \ interferences between a direct and a mixed
path\ will give measurements of the CPV parameters (\ref{a76}) equal to 
\begin{equation}
S_{f}=\sin 2\beta =\sin \left[ \arctan \left( 0.77\right) \right] =0.61\text{%
, }C_{f}=0\text{.}  \label{a85}
\end{equation}
Taking in Ref. \cite{5} a set of neutral final states $f$ \ = \{$J/\psi
K^{*0}$, $\phi K_{S}^{0}$, $K^{0}\pi ^{0}$\}, the reported experimental data 
give $S_{J/\psi K^{*0}}=0.60\pm 0.25$, $C_{J/\psi K^{*0}}=0.03\pm 0.1$, $%
S_{\phi K_{S}^{0}}=0.58\pm 0.12$, $C_{\phi K_{S}^{0}}=-0.09\pm 0.12$ and $%
S_{K^{0}\pi ^{0}}=0.64\pm 0.13$, $C_{K^{0}\pi ^{0}}=0.00\pm 0.08$. These
data are in good agreement with the gravity induced effect Eq. (\ref{a85}).
For other final states $f$, $S_{f}$ are centered around (\ref{a85}) but
deviate from this value, for example $S_{\psi K^{0}}=0.709\pm 0.017$, $%
C_{\psi K^{0}}=0.3\pm 1.0$ provides a canonical reference value above $%
5\sigma $. The fact that $\left\langle f\right| \mathcal{T}\left| \overline{B%
}^{0}\right\rangle /\left\langle f\right| \mathcal{T}\left|
B^{0}\right\rangle $ $\neq $ $1$ is the source of the dispersion of $\ S_{f}$%
, note also that the sign of $\left\langle f\right| CP\left| f\right\rangle $
is to be considered to analyze the sign of $S_{f}$.

A clear understanding of the $\sin 2\beta $ distribution, around $0.6-0.7$,
requires the adoption of the mass eigenstates (\ref{a71}, \ref{a72}) to
write down the data analysis protocols of the various experimental raw data
files and a precise evaluation of $\ \left\langle f\right| \mathcal{T}%
\left| \overline{B}^{0}\right\rangle /\left\langle f\right| \mathcal{T}%
\left| B^{0}\right\rangle $.

Let us adopt a second point of view. We will not consider the interpretation
of interferences experiments and, rather than addressing the issue of $\
\lambda _{f}$, we address directly the issue of $\ \beta $. We consider the
different mass eigenstates expansions on either CP or flavor eigenstates: (%
\ref{a11}, \ref{a12}, \ref{a73}, \ref{a74}) for the CPT one, and (\ref{a69}, 
\ref{a70}, \ref{a71}, \ref{a72}) for the gravity induced one. In order to
compare the usual eigenstates parametrization (\ref{a73}, \ref{a74}), based
on a single angle $\beta $, with the gravity induced mass eigenstates (\ref
{a71}, \ref{a72}), involving two parameters $\varkappa $ and $\chi $, we
must define $\beta $ through a \textit{gedanken} experiment providing $\exp
j\beta $ as a phase-convention-independent expression.

We consider the symmetric and complete combination 
\begin{equation}
\rho _{B}=\frac{\left\langle B^{0}\right. \left| B_{L}\right\rangle }{%
\left\langle \overline{B}^{0}\right. \left| B_{L}\right\rangle }\frac{%
\left\langle B^{0}\right. \left| B_{H}\right\rangle }{\left\langle \overline{%
B}^{0}\right. \left| B_{H}\right\rangle }\text{,}  \label{a86}
\end{equation}
which takes into account the four components at work in the CPV description.

This definition of $\beta $ through $\rho _{B}$ takes into account all
flavor and mass eigenstates but suffers from a lack of (unphysical) phase
compensation with respect to the flavor eigenstates. All measured
observables, independently of the interpretation of the measurement, are
combinations of phase-convention-independent quantities. We introduce the
coefficient $\varphi _{B}^{\prime }$ needed to provide a
phase-convention-independent observable associated with $\rho _{B}$%
\begin{equation}
\varphi _{B}^{\prime }=\frac{\left\langle \overline{B}^{0}\right. \left|
B_{2}\right\rangle \left\langle B_{2}\right. \left| B_{H}\right\rangle }{%
\left\langle B^{0}\right. \left| B_{1}\right\rangle \left\langle
B_{1}\right. \left| B_{L}\right\rangle }\frac{\left\langle \overline{B}%
^{0}\right. \left| B_{2}\right\rangle \left\langle B_{2}\right. \left|
B_{L}\right\rangle }{\left\langle B^{0}\right. \left| B_{1}\right\rangle
\left\langle B_{1}\right. \left| B_{H}\right\rangle }\text{,}  \label{a87}
\end{equation}
where we have chosen the two projection operators $\left| B_{1}\right\rangle
\left\langle B_{1}\right| $ and $\left| B_{2}\right\rangle \left\langle
B_{2}\right| $ because they commute with CP.

It can be checked that the product $\rho _{B}\varphi _{B}^{\prime }$ is
phase-convention-independent and thus can be measured in experiments.

If the usual CPT invariant parametrization of CPV effects is used (\ref{a11}%
, \ref{a12}, \ref{a73}, \ref{a74}), this rephasing factor $\varphi
_{B}^{\prime }$ changes nothing because it is equal to one 
\begin{eqnarray}
\rho _{B} &=&-\exp j4\beta \text{,}  \label{a88} \\
\varphi _{B}^{\prime } &=&1\text{,}  \label{a89}
\end{eqnarray}
and the product $\rho _{B}\varphi _{B}^{\prime }$ can be measured and
interpreted as -$\exp j4\beta .$

If CPV is gravity induced, we replace $\left| B_{H}\right\rangle $ and $%
\left| B_{L}\right\rangle $ with $\left| B_{H}^{\oplus }\right\rangle $ and $%
\left| B_{L}^{\oplus }\right\rangle $ given by (\ref{a69}, \ref{a70}, \ref
{a71}, \ref{a72}), and the very same observable is the product of the
following factors 
\begin{eqnarray}
\rho _{B}^{\oplus } &=&-1+O\left[ 10^{-6}\right] \text{,}  \label{a90} \\
\varphi _{B}^{\prime \oplus } &=&\frac{1+j\chi }{1-j\chi }=\exp \left(
2j\arctan \chi \right) \text{.}  \label{a91}
\end{eqnarray}
We conclude that, if gravity induced CPV is taken into account, the
measurement of the phase-convention-independent observable $\rho _{B}\varphi
_{B}^{\prime }$ on earth gives 
\begin{equation}
\rho _{B}^{\oplus }\varphi _{B}^{\prime \oplus }=-\exp \left( 2j\arctan \chi
\right) \text{,}  \label{a92}
\end{equation}
although if the measurement of the very same phase-convention-independent
observable $\rho _{B}\varphi _{B}^{\prime }$ is interpreted within the usual
CPT invariant framework it defines $\beta $ as 
\begin{equation}
\rho _{B}\varphi _{B}^{\prime }=-\exp j4\beta \text{.}  \label{a93}
\end{equation}
The conclusion of this $\rho _{B}$ \textit{gedanken} measurement with two
frameworks of interpretation is that $\arctan \chi =2\beta $ and 
\begin{equation}
\sin 2\beta =\sin \left[ \arctan \left( 0.77\right) \right] =0.61\text{.}
\label{a94}
\end{equation}
If we consider the entire Belle data samples in 2012 \cite{23}, plus the
entire BaBar data samples in 2009 \cite{24}, with about $10^{9}$ cumulated $%
B^{0}/\overline{B}^{0}$ pairs, the global average over the two data sets was
found, at that time, to be given by $\sin 2\beta =0.67\pm 0.02$ \cite{25}.
The mismatch between this experimental average and (\ref{a94}) is of
the order of \ 7\%.  Gravity induced CPV is thus a pertinent model to
interpret $B^{0}/\overline{B}^{0}$ experiments dedicated to CPV. This last
result complete the previous analysis of $K^{0}/\overline{K}^{0}$
experiments and confirm that the CKM matrix must be considered free from any
CPV phase far from any massive object.

\section{Gravity induced CPV in $D^{0}/\overline{D}^{0}$ and $B_{s}^{0}/%
\overline{B_{s}}^{0}$ experiments and conclusions}

The previous calculations on the impact of earth gravity on neutral mesons
oscillations can be extended to $D^{0}/\overline{D}^{0}$ $\sim \left( c%
\overline{u}\right) /\left( \overline{c}u\right) $ and $B_{s}^{0}/\overline{%
B_{s}}^{0}\sim \left( s\overline{b}\right) /\left( \overline{s}b\right) $.
The framework of analysis of the experimental data on $D^{0}/\overline{D}%
^{0} $ and $B_{s}^{0}/\overline{B_{s}}^{0}$ is similar to the methods
presented in section V, VI and VII. The parameters $m_{D}g\hbar /\delta
m_{D}^{2}c^{3}$ and $m_{B_{s}}g\hbar /\delta m_{B_{s}}^{2}c^{3}$ for both
mesons systems is very small so a type (\textit{i}) indirect violations will
be extremely difficult to observe. Type (\textit{ii}) and (\textit{iii}) can
be analyzed on the basis of gravity induced CPV, presented in section VI and
VII, but are also experimentally difficult to observe.

In any environment where a flavored neutral mesons $\left| M\right\rangle $,
with mass $m$, mass spliting $\delta m$ and Compton wavelength $\lambda _{C}$%
, experiences a gravity $\mathbf{g}$, i.e. in any curved space-time
environment, the amplitude of CP violation will be given by

\begin{equation}
\left( m/\delta m\right) ^{2}\left| \mathbf{g}\right| \lambda _{C}/c^{2}%
\text{.}  \label{a95}
\end{equation}
The first factor $m/\delta m$ is associated with electroweak and strong
interactions, the second one is the product of a (wave)length, an acceleration and $c$,
quantities related to geometry and space-time rather than to electroweak or
strong interactions. The proportionality to $\left| \mathbf{g}\right| $
indicate that this new CPV mechanism allows to set up cosmological evolution
models predicting the strong asymmetry between the abundance of matter and
the abundance anti-matter in our present universe \cite{6}.

Beside the problem of early baryogenesis, neutrinos oscillations near a
spherical massive object might be revisited to explore the impact of the
interplay between gravity and mixing.

The type (\textit{i}) CPV observed with $K^{0}/\overline{K}^{0}$ stems from
a gravity induced interplay between vertical quarks zitterbewegung
oscillations at the velocity of light on the one hand and the strangeness
oscillations $\left( \Delta S=2\right) $ on the other hand.

The type (\textit{ii}) small CPV observed with $K^{0}/\overline{K}^{0}$ is
associated with the CPT invariant modelisation of a gravity induced CPT
violation and is elucidated through a careful analysis of the rephasing
invariance of the observable $\eta _{00}$.

The large type (\textit{iii}) CPV observed with $B^{0}/\overline{B}^{0}$ is
associated with the CPT invariant modelisation of a gravity induced CPT
violation displaying a very small modulus and a significant phase $\beta$.

When the mesons are considered stables, the evolution is unitary and there
is no T violation, T violation stems from the modelisation of the transition
amplitudes $w_{f}$ \ in Eq. (\ref{a1}) as irreversible decays in Eq. (\ref
{a2}) within the framework of the WW approximation \cite{10}.

The very large type (\textit{iv}) CPV observed in our universe, namely its
baryon-antibaryon asymmetry, remains an open issue within the KM framework
of interpretation, although gravity induced CPV displays the potential to
set up cosmological evolution models in agreement with the present state of
our universe.

We have demonstrated that gravity induced CPV allows to predict three
experimental CPV parameters ($\varepsilon ,\varepsilon ^{\prime },\beta $)
and appears to provide the potential to explain the baryon asymmetry of the
universe as its amplitude is linear with respect to the strength of gravity. 

This set of new results was obtained within the canonical framework of
quantum mechanics, on earth, without any speculative assumption on new
coupling, or new field, or new physics. From this clear 
convergence of results, we can conclude that a CKM matrix free of CPV phase
is to be considered as the core of the SM in a flat Lorentzian environment
and earth's gravity is the sole source of $\ \varepsilon $, $\varepsilon
^{\prime }$ and $\beta $ CPV effects in $K^{0}/\overline{K}^{0}$ and $B^{0}/%
\overline{B}^{0}$ experiments.


\begin{thebibliography}{500}

\bibitem{1}  J. H. Christenson, J. W. Cronin, V. L. Fitch, and R. Turlay,
Evidence for the $2\pi $ decay of the $K_{2}^{0}$ meson, Phys. Rev. Lett. 
{\bf 13}, 138 (1964).

\bibitem{2}  M. Kobayashi and T. Maskawa, CP-violation in the renormalizable
theory of weak interaction, Progress of Theoretical Physics {\bf 49}, 652
(1973).

\bibitem{3}  N. Cabibbo, Unitary symmetry and leptonic decays, Phys. Rev.
Lett. {\bf 10}, 531 (1963).

\bibitem{4}  T. D. Lee, {\it Particle physics and introduction to field
theory} (Harwood Academic, New york, 1981).

\bibitem{5}  P. D. Group, Review of particle physics, Phys. Rev. D {\bf 110}, 030001 (2024).

\bibitem{6}  A. D. Sakharov, Violation of CP invariance, C asymmetry, and
baryon asymmetry of the universe, Pisma Zh. Eksp. Teor. Fiz. 5, {\bf 32}
(1967).

\bibitem{7}  T. D. Lee, R. Oehme, and C. N. Yang, Remarks on possible
noninvariance under time reversal and charge conjugation, Phys. Rev. {\bf 106}, 340 (1957).

\bibitem{8}  T. T. Wu and C. N. Yang, Phenomenological analysis of violation
of CP invariance in decay of $K^{0}$ and $\overline{K}^{0}$, Phys. Rev.
Lett. {\bf 13}, 380 (1964).

\bibitem{9}  J. M. Rax, Zitterbewegung CP violation in a Schwarzschild spacetime, 
arXiv 2403.07970 (03-2024).

\bibitem{10}  V. Weisskopf and E. Wigner, Berechnung der naturlichen
linienbreite auf grund der diracschen lichttheorie, Zeitschrift fur Physik 
{\bf 63}, 54 (1930).

\bibitem{11}  J. Bjorken and S. Drell, {\it Relativistic Quantum Mechanics}
(McGraw-Hill, New york, 1964).

\bibitem{12} J. M. Rax, Gravity induced CP violation, arXiv 2405.17317 (05-2024).

\bibitem{13} J. M. Rax, {\it Mecanique Analytique} (Dunod Sciences-Sup,
Paris, 2020).

\bibitem{14}  E. Fischbach, Test of general relativity at the quantum level,
in {\it Proceedings of the NATO Advanced Study Institute on Cosmology and
Gravitation}, edited by P. Bergmann and V. de Sabbata, NATO Scientific
Affairs Division (Plenum Press, New york and London, 1980) pp. 359-373.

\bibitem{15}  G. Chardin and J. M. Rax, CP violation. a matter of
(anti)gravity?, Phys. Lett. B {\bf 282}, 256 (1992).

\bibitem{16}  J. Bell and J. Steinberger, in {\it Proceedings of the Oxford
International Conference on Elementary Particles 1965}, edited by R. G.
Moorhouse, A. E. Taylor, and T. R. Walsh (Rutherford High Energy Laboratory,
1966) p. 195.

\bibitem{17}  H. Burkhardt et al. (NA31), First evidence for direct CP
violation, Phys. Lett. B {\bf 206}, 169 (1988).

\bibitem{18}  V. Fanti et al. (NA48), A new measurement of direct CP
violation in two pion decays of the neutral kaon, Phys. Lett. B {\bf 465}, 335 (1999).

\bibitem{19}  A. Alavi-Harati et al. (KteV), Observation of direct CP
violation in $K_{S,L}\rightarrow \pi \pi $ decays, Phys. Rev. Lett. {\bf 83}, 22 (1999).

\bibitem{20}  B. Aubert et al. (BaBar), Observation of CP violation in the $%
B_{0}$ meson system, Phys. Rev. Lett. {\bf 87}, 091801 (2001).

\bibitem{21}  K. Abe et al. (Belle), Observation of large CP violation in
the neutral $B$ meson system, Phys. Rev. Lett. {\bf 87}, 091802 (2001).

\bibitem{22}  B. Aubert et al. (BaBar), Improved measurement of CP violation
in neutral $B$ decays to $c\overline{c}s$, Phys. Rev. Lett. {\bf 99}, 171803
(2007).

\bibitem{23}  I. Adachi et al. (Belle), Precise measurement of the CP
violation parameter, Phys. Rev. Lett. {\bf 108}, 171802 (2012).

\bibitem{24}  B. Aubert et al. (BaBar), Measurement of time-dependent
asymmetry in decays, Phys. Rev. D {\bf 79}, 072009 (2009).

\bibitem{25}  Y. Nir and V. Vagnoni, CP violation in B decays, Compte
Rendus Physique {\bf 21}, 1 (2020).

\end{thebibliography}
\end{document}